\documentclass[10pt, letter, twocolumn]{IEEEtran}

\usepackage{graphicx}
\usepackage{epsfig}
\usepackage{algorithm}
\usepackage{algorithmic}
\usepackage{ifmtarg}
\usepackage{caption}
\usepackage{footnote}
\usepackage{cite}
\usepackage{amssymb}
\usepackage{amsthm}
\usepackage[fleqn]{amsmath}
\usepackage{amsmath}
\usepackage{mathtools}
\usepackage{amsfonts}
\usepackage{epstopdf}
\usepackage{enumitem}
\usepackage{subfigure}
\usepackage{multirow}
\usepackage{color}
\usepackage[utf8]{inputenc}
\usepackage[english]{babel}

\usepackage[font=normalsize]{caption}
\theoremstyle{definition}
\newtheorem{defn}{Definition}

\usepackage[]{graphicx}
\usepackage{subfigmat}
\usepackage{changepage}

\begin{document}
\title{Multi-band RF Energy and Spectrum Harvesting in Cognitive Radio Networks}
\author{
\IEEEauthorblockN{Ahmad Alsharoa, Nathan M. Neihart, Sang W. Kim, Ahmed E. Kamal}\\
\IEEEauthorblockA{Iowa State University (ISU), Ames, Iowa, United States, \\ Email: \{alsharoa, neihart, swkim, kamal\}@iastate.edu}\vspace{-1cm}}
\maketitle
\thispagestyle{empty}
\pagestyle{empty}

\begin{abstract}
This paper investigates a multi-band harvesting (EH) schemes under cognitive radio interweave framework.
All secondary users are considered as EH nodes that are allowed to harvest energy from multiple bands of Radio Frequency (RF) sources.
A win-win framework is proposed, where SUs can sense the spectrum to determine
whether the spectrum is busy, and hence they may harvest from RF energy, or if it is idle, and hence they can use it for transmission. Only a subset of the SUs can sense in order to reduce sensing energy, and then machine learning is used to characterize areas of harvesting and spectrum usage.
We formulate an optimization problem that jointly optimize number of sensing samples and sensing threshold in order to minimize the sensing time and hence maximize the amount of energy harvested.
A near optimal solution is proposed using Geometric Programming (GP) to optimally solve the problem in a time-slotted period. Finally, an energy efficient approach based on multi-class Support Vector Machine (SVM) is proposed by involving only training SUs instead of all SUs.
\end{abstract}

\vspace{-.3cm}
\section{Introduction}
With high demands of future generation wireless communication networks, researchers have focused on proposing efficient and smart solutions for energy and spectrum deficit problems. Prospective demands of future generation wireless networks will require achieving around 1000 times higher data rates and 10 times lower round-trip latency~\cite{andrews2016what}.
The profile ratio of the Internet of Things (IoT) and the increasing number of mobile devices, are contributing to these requirements.
This is also coupled with IoT requirements of low energy communication.

Cognitive radio is proposed as a novel solution to solve the spectrum deficit problems. The basic idea of cognitive radio is that unlicensed or Secondary Users (SUs) are allowed to utilize the spectrum of licensed users which are also known as primary users (PUs) in an opportunistic manner~\cite{CR2,CR3}. 

Energy harvesting (EH) is considered as one of the most robust methods to mitigate the energy consumption problem and perpetuate the lifetime and sustainability of wireless systems where around $30\%$ of the energy expenditure of mobile devices is caused by wireless networking~\cite{pollin2008meera}. Radio frequency (RF) EH, which is also known as wireless energy transfer, has been introduced as an effective harvesting technique where energy is collected from RF sources generated by other neighbor devices. nights~\cite{7010878}.

Most of the previous works on RF EH focused on harvesting from a single band, which limit the amount of the harvested energy \cite{CA}.
Recently, few circuit designs that use a single antenna~\cite{single2,single} 
or multiple antennas~\cite{multiple,multiple2} to harvest from multiple band have been introduced in the literature, where the multi-band concept increases the amount of harvested energy to the area ratio.
For instance, in~\cite{single2}, a dual-band circuit using single antenna has been designed to harvest from 1.8GHz and 2.1GHz frequency bands. Another multi-band circuit designating one wide antenna has been proposed in~\cite{single}, where it achieves a good
efficiency around 40\% for -15 dBm received power. While, in~\cite{multiple2} Keyrouz \textit{et.al} designed multi-band RF with triple antennas using cascaded rectifying stages. With -15 dBm received power, their proposed design achieves efficiency around \{45, 46, 25\}\% for \{0.9, 1.8, 2.45\}~GHz, respectively. In~\cite{multiple}, the authors used multiple antennas for harvesting and showed that the amount of harvested energy can be increased by harvesting from multiple bands instead of a single band.

In this paper we investigate the classification problem of SUs, where SUs perform sensing of the spectrum in order to determine 1) harvesting geographical regions in which SUs in that region can harvest RF energy, and
2) communication geographical regions in which SUs in that region can use the spectrum without causing interference to the PUs.
Furthermore, we propose an efficient solution based on multi-class Support Vector Machine (SVM)~\cite{Vapnik}. This is therefore a win-win strategy for most of SUs because of the result of sensing they either use the spectrum or harvest energy from RF signals.
The main contributions of this paper are summarized as follows
\begin{itemize}
  \item Investigating RF multi-band EH under cognitive radio framework, where SUs use the harvested energy to perform sensing. For this reason, a time switching protocol is adopted where SUs switch over time between sensing and EH.
  \item Analyzing RF multi-band EH by using a wide antenna with multiple rectifying stages.
  \item Formulating a time-slotted optimization problem that aims to classify the belonging regions of SUs and minimize the sensing time.
  \item Optimizing the number of sensing samples and sensing threshold for each RF band and finding the SUs that can harvest from RF multi-band signals.
  \item Proposing an energy efficient solution based on multi-class SVM to determine the harvesting and communication regions of SUs.
\end{itemize}


\section{System Model}\label{system}
In this work, we investigate a time-slotted system of a finite period of time slots $t=1,..,\Gamma$, each of equal duration $T$.

\subsection{Network Model}
We consider a cognitive radio network consisting of $U$ PUs and $M$ time synchronous EH SUs using RF signals. Each PU transmits its signal over one frequency band. Without loss of generality, we assume that each SU can sense energy from RF signals from $\mathbb{B}$ bands using multiple rectifying stages.

\begin{figure}[h!]
  \centerline{\includegraphics[width=2.05in]{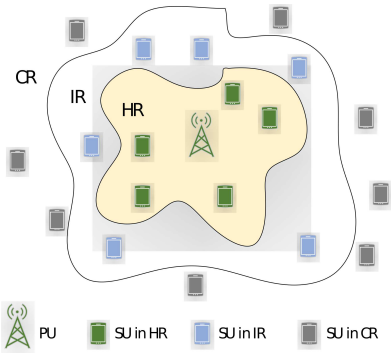}}
   \caption{System model}\label{Proposal_systemmodel}
\end{figure}

Based on the received power at the SUs, each SU can be in one or more of three regions as shown in Fig~\ref{Proposal_systemmodel}. The first one is Harvesting Region (HR), where the received power at the SU is greater than a specific sensitivity threshold $P_\text{th}$ which will allow energy to be harvested~\cite{sen_th}. This threshold is due to the limitation of RF-to-DC circuit sensitivity. The second region called Communication Region (CR) contains the SUs that can not sense the PU channel and therefore, the channel maybe used for communication. The middle one is Inactive Region (IR) with no harvesting and no communication, where all the SUs can sense the PU signal but the received power is not enough to activate the harvesting circuit (i.e., the received power is less than $P_\text{th}$). Note that, in IR, underlay cognitive radio can be used where the SUs are allowed to transmit their signals under a certain primary interference threshold~\cite{ETTalsharoa}, but the use of the spectrum is beyond the scope of this paper.
Note that it is possible for SU $m$ to be in more that one region at the same time (e.g., SU $m$ can be in HR for bands $b=\{1,3\}$, in IR for band $b=2$, and in CR for bands $b=\{4,5\}$).
In this paper, we study the sensing performance of multi-band system in order to classify the SUs according to the three regions above. 
\subsection{Channel Model}
We denote by $h^b_{u,m,t}$ the channel gain during time slot $t$ between PU $u$ and SU $m$ over band $b$ and is given by $h^b_{u,m,t}=G_t G_r \left( \frac{c}{4 \pi d_{u,m} f^b} \right)^2$,
where $d_{u,m}$ is the Euclidean distance between the PU and SU~$m$ in meter. $c$ and $f^b$ are the speed of light in m/s and transmission frequency in Hz, respectively.
Without loss of generality, all channel gains are assumed to be constant during the whole time slot of $T$ sec, i.e., during sensing and/or harvesting times.

\subsection{Spectrum Sensing}
In this paper an \textit{energy detection} method is used for spectrum sensing, where energy detectors (EDs) have been exploited as an efficient technique in the literature due to their simplicity, compatibility with any signal type, and low implementation complexity \cite{axell2012spectrum}. A binary hypothesis testing can be built to determine the PU status as follows:
to detect PU channel, ED of SUs measure the received signal energy for a number of samples, $\vartheta^b_{m,t}$, and compares the cumulative energy to a threshold $\varepsilon^b_{m,t}$ for each band $b$ as $\sum\limits_{v=1}^{\vartheta^b_{m,t}} (y^b_{m,t}(v))^2 \underset{\mathcal{H}_0}{\overset{\mathcal{H}_1}{\gtrless}} \varepsilon^b_{m,t}$,
where $(y^b_{m,t}(v))^2$ is the energy measured on sample $v$. $\mathcal{H}_0$ represents the absence of the primary signal, i.e., the received baseband complex signal contains only Additive White Gaussian Noise (AWGN) and $\mathcal{H}_1$ represents the presence of the primary signal.
For a large enough number of samples $\bar{\vartheta}$, (i.e., $\vartheta^b_{m,t} \geq \bar{\vartheta}$) and normalized noise variance, the probability of false alarm, $F^b_{m,t}$, and  probability of detection, $D^b_{m,t}$, can be respectively expressed as~\cite{formulas}
\begin{align}
F^b_{m,t}&= \mathcal{Q}\left[\left(\varepsilon^b_{m,t} - 1\right)\sqrt{\vartheta^b_{m,t}}\textbf{ }\right] \label{eq:falseAlarm} \\
D^b_{m,t}&=\mathcal{Q} \left[\frac{  \left(\varepsilon^b_{m,t} - \gamma^b_{m,t} -1\right) \sqrt{\vartheta^b_{m,t}}}{\gamma^b_{m,t}+1}\textbf{ }\right] \label{eq:Detection}
\end{align}
where $\gamma^b_{m,t}$ and $\mathcal{Q}\left[\cdot\right]$ denote the signal-to-noise ratio (SNR) of SU$_m$ received at band $b$ and the Q-function, respectively.

\subsection{Energy Harvesting}
Let us assume that $\delta^b_{m,t}$ is the harvesting sensitivity indicator function which is equal to 1 (i.e., SU $m$ over band $b$ is in HR) if the average received power over band $b$ greater than $P_\text{th}$ and 0 otherwise, and given as follows
\begin{equation}\label{delta_eq}
\small
\delta^b_{m,t}=
\left\{
   \begin{array}{ll}
   1, & \left( \frac{\sum\limits_{v=1}^{\vartheta^b_{m,t}} P^b_t(v) h^b_{u,m,t}}{\vartheta^b_{m,t}} \geq P_\text{th} \right) / \mathcal{H}_1,\\
   0, & \hbox{otherwise}
   \end{array}
\right.
\normalsize
\end{equation}
where $P^b_t(v)$ is the transmit power of PU on band $b$ during time slot $t$ of sample $v$. The condition $(\sum_v^{\vartheta^b_{m,t}} P^b_t(v) h^b_{u,m,t})/\vartheta^b_{m,t}\geq P_\text{th}$ is to ensure that the SUs is in HR.

\begin{figure}[h!]
  \centerline{\includegraphics[width=3in]{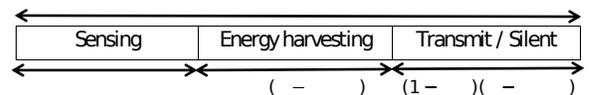}}
   \caption{Energy harvesting scheme.}\label{Proposal_systemmodel22}
\end{figure}

Let us assume that $T_s$ is the sensing time for one sample. Hence, During $(\vartheta^b_{m,t} T_s)$ seconds, SU $m$ senses PU channel over band $b$ to determine the belonging region (i.e., HR, IR, CR). For the remaining time, $\delta^b_{m,t} (T-\vartheta^b_{m,t} T_s)$ seconds, SU~$m$ can harvest if it belongs to the HR. Otherwise, during $(1-\delta^b_{m,t}) (T-\vartheta^b_{m,t} T_s)$ seconds, SU $m$ can transmit its data if it belongs to CR, or keep silent if it belongs to IR as shown in Fig.~\ref{Proposal_systemmodel22}.
Notice that, the choice of this $\vartheta^b_{m,t}$ affects the performance of the SUs. In fact, decreasing $\vartheta^b_{m,t}$ allows SU $m$ to harvest more energy that can be used for future sensing or transmission. However, this will reduce the allocated time to perform sensing and vice versa. Therefore, an optimal choice of $\vartheta^b_{m,t}, \forall m=1,..,M$ is required in order to enhance the overall system performance.

\section{Multi-band Energy Harvesting Scheme}\label{multibands}
\begin{figure}[h!]
  \centerline{\includegraphics[width=3in]{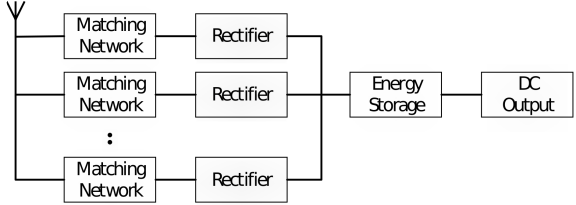}}
   \caption{Multi-bands energy harvesting scheme}\label{single}
\end{figure}

Several RF harvester schemes can be used for multi-band EH. The main difference lies in the RF bandpass filter design and number of antennas. 
In this scheme, we consider that each SU is equipped with a single wideband antenna feeding parallel rectifier circuits at dedicated bands as shown in Fig.~\ref{single}.
where the matching network is designed to match each parallel rectifier at the dedicated frequencies.
The total consumed sensing energy of SU $m$ during time slot $t$, denoted by $S_{m,t}$ can be expressed~as
\begin{equation}\label{Se}
S_{m,t}= \sum\limits_{b=1}^\mathbb{B}  \vartheta^b_{m,t} T_s P_s,
\end{equation}
where $P_s$ is the sensing circuit power consumption per sample.

The total harvested energy of SU $m$ during time slot $t$, denoted by $H_{m,t}$ can be written~as
\begin{equation}\label{Eh}
H_{m,t}= \sum\limits_{b=1}^\mathbb{B} \delta^b_{m,t} \eta^b_{m,t} \left[(T-\vartheta^b_{m,t} T_s) P^b_t h^b_{u,m,t}  \right],
\end{equation}
where $\eta^b_{m,t}$ is the RF-to-DC efficiency and depends on rectifying technology and operated frequency.
The stored energy at the end of time slot $t$ at SU~$m$, denoted by $B_{m,t}$, is given as
\begin{equation}\label{Battery}
B_{m,t}= \max(B_{m,t-1}+ H_{m,t}- S_{m,t},0),
\end{equation}
where $\max(x,0)$ takes the maximum value between $x$ and 0.

\section{Problem Formulation}\label{formulation}
In this section, the optimization problem that minimizes the total sensing energy of the SUs is formulated by taking into consideration the detection and false alarm probabilities constraints in addition to SUs battery constraints.


The false alarm and detection probabilities need to satisfy the following conditions:
\begin{equation}\label{Pm_con0}
F^b_{m,t} \leq \bar{F}, \quad D^b_{m,t}\geq  \bar{D}, \quad \forall m, \forall t, \forall b,
\end{equation}
where $\bar{F}$ and $\bar{D}$ are the false alarm and detection thresholds. By substituting \eqref{eq:falseAlarm} and \eqref{eq:Detection} in the above equations and with simple manipulations \eqref{Pm_con0} can be written, respectively,~as
\small
\begin{equation}
\frac{ \varepsilon^b_{m,t} (\vartheta^b_{m,t})^\frac{1}{2}}{\mathcal{Q}^{-1} \left[ \bar{F} \right] +\vartheta^b_{m,t}  } \leq 1, \forall m, \forall t, \forall b,
\end{equation}
\begin{equation}
\frac{ (\gamma^b_{m,t}+1)\mathcal{Q}^{-1} \left[ \bar{D} \right] + (\gamma^b_{m,t}+1) (\vartheta^b_{m,t})^\frac{1}{2}}{\varepsilon^b_{m,t} (\vartheta^b_{m,t})^\frac{1}{2}}\leq 1,\forall m, \forall t, \forall b.
\end{equation}
\normalsize
Therefore, our optimization problem can be formulated as
\begin{align}
&\hspace{-0.5cm}\underset{\varepsilon^b_{m,t} \geq 0, \vartheta^b_{m,t}\in \mathbb{R^+}}{\text{minimize}} \quad  \quad \sum \limits_{m=1}^{M}  \sum\limits_{b=1}^\mathbb{B}  \vartheta^b_{m,t} T_s P_s     \label{of}\\
&\hspace{-0.5cm}\text{subject to:}\nonumber\\
&\hspace{-0.5cm} \frac{ \varepsilon^b_{m,t} (\vartheta^b_{m,t})^\frac{1}{2}}{\mathcal{Q}^{-1} \left[ \bar{F} \right] +\vartheta^b_{m,t}  } \leq 1, \quad \forall m, \forall b,  \label{Pm_con}  \\
&\hspace{-0.5cm} \frac{ (\gamma^b_{m,t}+1)\mathcal{Q}^{-1} \left[ \bar{D} \right] + (\gamma^b_{m,t}+1) (\vartheta^b_{m,t})^\frac{1}{2}}{\varepsilon^b_{m,t} (\vartheta^b_{m,t})^\frac{1}{2}} \leq 1, \quad \forall m, \forall b, \label{Pd_con}  \\
&\hspace{-0.5cm} \sum\limits_{b=1}^\mathbb{B}  \vartheta^b_{m,t} T_s P_s \leq B_{m,t-1},  \quad \forall m, \label{Sm_con}  \\
&\hspace{-0.5cm} \bar{\vartheta} \leq \vartheta^b_{m,t} \leq \frac{T}{T_s}, \quad \forall m, \forall b, \label{Nm_con}
\end{align}
where, constraints \eqref{Pm_con} and \eqref{Pd_con} represent the false alarm and detection probabilities constraints, respectively. Constraint \eqref{Sm_con} represents battery causality constraint.
Constraint \eqref{Nm_con} represents the number of samples constraint.

\section{Joint Optimization Solution based on Geometric Programming}\label{jsolutions}

Notice that the formulated optimization problem in \eqref{of}-\eqref{Nm_con} is a mixed-integer non-linear problem (MINP). In order to simplify the problem, we propose to proceed with a joint-optimization approach where we optimize $\vartheta^b_{m,t}$ and $\varepsilon^b_{m,t}$ jointly using Geometric Programming (GP). To do this, we firstly relax the variable $\vartheta^b_{m,t}$ and make it continuous. After obtaining an optimal value of  $\vartheta^b_{m,t}$, we can obtain the closest upper integer value, which does not negatively affect the system performance.
Furthermore, we apply a successive convex approximation (SCA) approach to transform the non-convex problem into a sequence of relaxed convex subproblems~\cite{alsharoaGP}.

\subsection{Geometric Programming Method}
\label{GPmethod}
%
The standard form of GP is defined as the minimization of a posynomial function subject to inequality posynomial constraints and equality monomial constraints as given below:
\begin{align}
&\underset{\boldsymbol{z}}{\text{minimize}} \quad  \quad f_0(z)\\
&\text{subject to:}\nonumber\\
& f_l(z) \leq 1, \quad \forall l=1,\cdots,L,\label{posyinomialc}\\
&\tilde{f}_{\tilde{l}}(z) = 1, \quad \forall \tilde{l}=1,\cdots,\tilde{L},\label{moninomialc}
 \end{align}
where $f_l(z)$, $l=0,\cdots,L$, are posynomials and $\tilde{f}_{\tilde{l}}(z)$, $\tilde{l}=1,\cdots,\tilde{L}$ are monomials. A monomial is defined as a function $f:$ \textbf{R}$^n_{++}$ $\rightarrow$ \textbf{R} as follows:$f(z)=\check{s}\ z_1^{s_1} z_2^{s_2}...\ z_N^{s_N}$,
where the multiplicative constant $\check{s} \geq 0$, and the exponential constants $s_n$ $\in$ \textbf{R}, $n=1,...,N$. A posynomial is a non-negative sum of monomials.

In general, GP in its standard form is a non-convex optimization problem, because posynomials and monomials functions are non-convex functions. However, with a logarithmic change of the variables, objective function, and constraint functions, the optimization problem can be turned into an equivalent convex form using the property that the logarithmic sum of exponential functions is convex (see~\cite{boyd} for more details).
In order to convert the optimization problem formulated in~\eqref{of}-\eqref{Nm_con} to a GP standard form, we propose to apply approximation for only constraint~\eqref{Pm_con}. The single condensation method is employed to convert this constraint to posynomial as described below:
\begin{defn}
The single condensation method for GP involves upper bounds on the ratio of a posynomial over a posynomial. It is applied to approximate a denominator posynomial $g(z)$ to a monomial function, denoted by $\tilde{g}(z)$ and leaving the numerator as a posynomial, using the arithmetic-geometric mean inequality as a lower bound~\cite{alsharoaGP}. Given the value of $z$ at the iteration $r-1$ of the SCA $z^{(r-1)}$, the posynomial $g$ that, by definition, has the form $g(z)\triangleq\sum^K_{k=1}\mu_k(z)$, where $\mu_k(z)$ are monomials, can be approximated as:
\begin{equation}\label{bound}
 g(z) \geq \tilde{g}(z)=\prod^K_{k=1} \left(\frac{\mu_k(z)}{\tilde{\mu}_k(z^{(r-1)})} \right)^{\tilde{\mu}_k(z^{(r-1)})},
 \end{equation}
where $\tilde{\mu}_k(z^{(r-1)})=\frac{\mu_k(z^{(r-1)})}{g(z^{(r-1)})}$. $K$ corresponds to the total number of monomials in $g(z)$.
\end{defn}
It can be noticed that the nominator and denominator of~\eqref{Pm_con} are posynomials, however, the ratio is not necessary a posynomial. Therefore, in order to convert constraint~\eqref{Pm_con} to a posynomial, we propose to apply the single condensation method given in Definition 1 to approximate the denominator posynomial to a monomial function, where in this case $K=2$, therefor constraint~\eqref{Pm_con} can be given as follows
\begin{equation}\label{Pm_con_mod}
\frac{ \varepsilon^b_{m,t} (\vartheta^b_{m,t})^\frac{1}{2}}{\tilde{g}(\vartheta^b_{m,t})} \leq 1, \quad \forall m, \forall b,
\end{equation}
It can be seen that constraint~\eqref{Pm_con_mod} is now a posynomial because a posynomial over a monomial is a posynomial. The convergence proof is given in \cite{C_GP}.
By considering the approximations of~\eqref{Pm_con_mod}, we can formulate the GP approximated subproblem at the $i^\text{th}$ iteration of the SCA as follows:
\begin{align}
&\hspace{-0.5cm}\underset{\varepsilon^b_{m,t}, \vartheta^b_{m,t}\geq 0  }{\text{minimize}} \quad  \quad \sum \limits_{m=1}^{M}  \sum\limits_{b=1}^\mathbb{B}  \vartheta^b_{m,t} T_s P_s     \label{ofi}\\
&\hspace{-0.5cm}\text{subject to:}\nonumber\\
&\hspace{-0.5cm} \frac{ \varepsilon^b_{m,t} (\vartheta^b_{m,t})^\frac{1}{2}}{\tilde{g}(\vartheta^b_{m,t})} \leq 1, \quad \forall m, \forall b,  \label{Pm_coni}  \\
&\hspace{-0.5cm} \frac{ (\gamma^b_{m,t}+1)\mathcal{Q}^{-1} \left[ \bar{D} \right] + (\gamma^b_{m,t}+1) (\vartheta^b_{m,t})^\frac{1}{2}}{\varepsilon^b_{m,t} (\vartheta^b_{m,t})^\frac{1}{2}} \leq 1, \quad \forall m, \forall b, \label{Pd_coni}  \\
&\hspace{-0.5cm} \frac{\sum\limits_{b=1}^\mathbb{B}  \vartheta^b_{m,t} T_s P_s}{B_{m,t-1}} \leq 1  \quad \forall m, \label{Sm_coni}  \\
&\hspace{-0.5cm}  \frac{T_s}{T} \vartheta^b_{m,t} \leq 1, \quad \forall m, \forall b, \label{Nm_conii} \\
&\hspace{-0.5cm} \frac{\bar{\vartheta}}{\vartheta^b_{m,t}} \leq 1, \quad \forall m, \forall b, \label{Nm_conii}
\end{align}

We solve the problem in an online fashion (i.e., time slot by time slot) for $t=1,..,\Gamma$.
Two solutions are proposed: The first one namely "\textit{all sensing}" solution where all secondary users perform sensing to determine to which regions they belong to.
This is an exhaustive solution and consumes a lot of energy but gives an accurate characterization or HR, IR, and CR.
The second solution is based on SVM where training of some SUs make them perform sensing in order to characterize the belonging region using multi-class classification. The latter solution is proposed to save some sensing energy by allowing only training SUs to perform sensing while the others keep quiet until they know their classification, however, the \textit{all sensing} solution has a superior performance over the SVM solution in terms of accuracy.

\subsection{Solution 1: All Sensing Solution}
The optimization problem formulated in \eqref{ofi}-\eqref{Nm_conii} can be solved optimally at each iteration of the SCA as given in Algorithm~\ref{SCAAlgorithm} using Gurobi/CVX interface \cite{Gurobi}. Each GP in the iteration loop (line 3-7) tries to improve the accuracy of the approximations to a particular minimum in the original feasible region. This is performed until no improvement in the objective function ($\chi$) is made. A parameter, $\varpi\rightarrow 0$, is introduced to control the accuracy of the algorithm convergence as follows: $|\chi^{(r+1)}-\chi^{(r)}|\leq \varpi$.
\begin{algorithm}[h!]
\caption{SCA Algorithm}
\label{SCAAlgorithm}
\small
\begin{algorithmic}[1]
\FOR {$t=1 \cdots \Gamma$}
\STATE r=1.
\STATE Select a feasible initial value of $\boldsymbol{z}^{(r)}= [\varepsilon^b_{m,t}, \vartheta^b_{m,t}], \forall m, \forall b$.
\REPEAT
\STATE r=r+1.
\STATE Approximate the denominators using the arithmetic-geometric mean as indicated in~\eqref{bound} using $\boldsymbol{z}^{(r-1)}$.
\STATE Solve the optimization problem using the interior-point method to determine the new approximated solution $\boldsymbol{z}^{(r)}= [\varepsilon^b_{m,t}, \vartheta^b_{m,t}], \forall m, \forall b$.
\UNTIL Convergence ($|\chi^{(r+1)}-\chi^{(r)}|\leq \varpi$).
\STATE Compute the corresponding harvested energy and battery level using~\eqref{Eh} and~\eqref{Battery}, respectively.
\ENDFOR
\end{algorithmic}
\normalsize
\end{algorithm}

\subsection{Solution 2: SVM Solution}
Several binary classification approachs can be extended to handle the SVM multi-class.
Two main types of approaches have been proposed for multi-class SVM by combining several binary classifications namely One-vs-All (OVA) and One-vs-One (OVO). The OVO approach considering all classes optimization in one formulation, therefore, it leads to more accurate results than OVA approach. However, OVO approach is more computational expensive than the OVA approach because it leads to larger scale optimization problem~\cite{mSVMS}.
In this paper we propose to use OVO approach as suggested in~\cite{mSVMS}.
This approach constructs $Y(Y-1)/2$ SVM classifiers where each classifier trains on data from two classes, where $Y$ is the number of classes.
For training data from class $i$ and class $j$, we solve the following binary classification problem

\begin{align}
&\hspace{-0.5cm}\underset{w^{ij}, \beta^{ij}, \zeta\tau^{ij}\geq 0}{\text{minimize}} \quad  \quad \frac{1}{2} (w^{ij})^T w^{ij} + \bar{\zeta}  \sum\limits_\tau \zeta_\tau^{ij} \label{ofSVM}\\
&\hspace{-0.5cm}\text{subject to:}\nonumber\\
&\hspace{-0.5cm} (w^{ij})^T \phi(x_\tau)+ \beta^{ij} \geq 1-\zeta_\tau^{ij}, \quad \text{if} \quad Y_\tau=i,  \label{c1SVM}  \\
&\hspace{-0.5cm} (w^{ij})^T \phi(x_\tau)+ \beta^{ij} \leq -1+\zeta_\tau^{ij}, \quad \text{if} \quad Y_\tau=j,  \label{c2SVM},
\end{align}
where the training data $x$ are mapped into a higher dimensional space of function $\phi(x)$. Minimizing $\frac{1}{2} (w^{ij})^T w^{ij}$ is equivalent to minimizing $2/||w^{i,j}||$, the margin between two groups of data.
The penalty term $\bar{\zeta}  \sum\limits_\tau \zeta_\tau^{ij}$ can reduce the number of training errors when the data is not linearly separable.
The work in \cite{Friedman} suggests to use a voting approach called Max-Wins approach, hence, if sign$( (w^{ij})^T \phi(x)+ \beta^{ij})$ gives $x$ is in class $i$, then the vote for class $i$ increases by one to this class. Otherwise, class $j$ is increased by one. Finally, $x$ can be predicted as the class with the largest vote.
Practically we solve the quadratic program formulated in~\eqref{ofSVM}-\eqref{c2SVM} using dual method where number of variables is equal to the number of the training data in two classes~\cite{boyd}.

\section{Simulation Results}\label{simulation}

\begin{figure*}
\vspace{-2mm}
\begin{tabular}{l}
\begin{minipage}[b]{0.9775\textwidth}
\begin{center}
\begin{subfigmatrix}{3}
   \subfigure[Band 1, All sensing]{\includegraphics[width=1.9in]{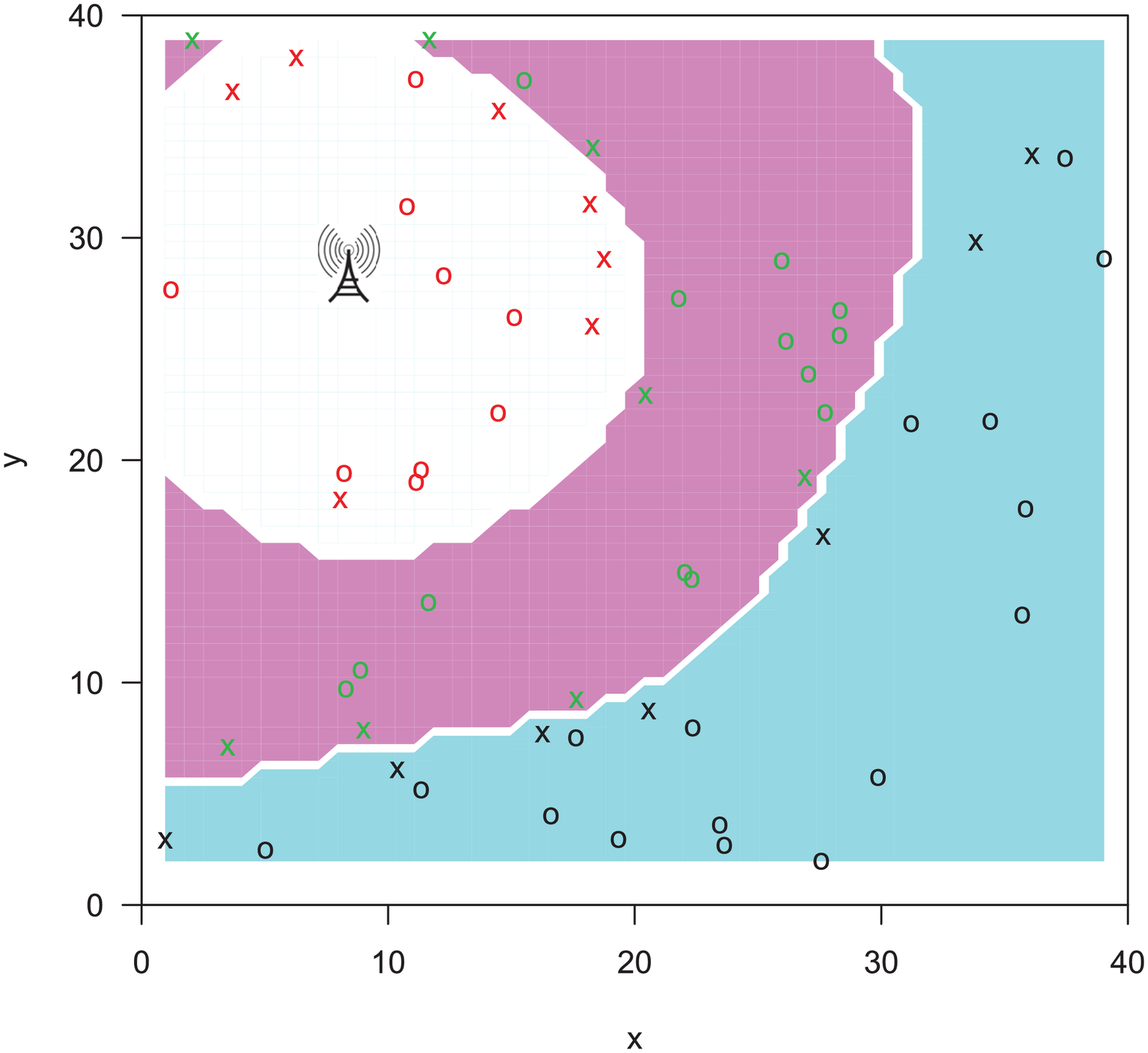}}
   \hspace{-3cm}\subfigure[Band 2, All sensing]{\includegraphics[width=1.9in]{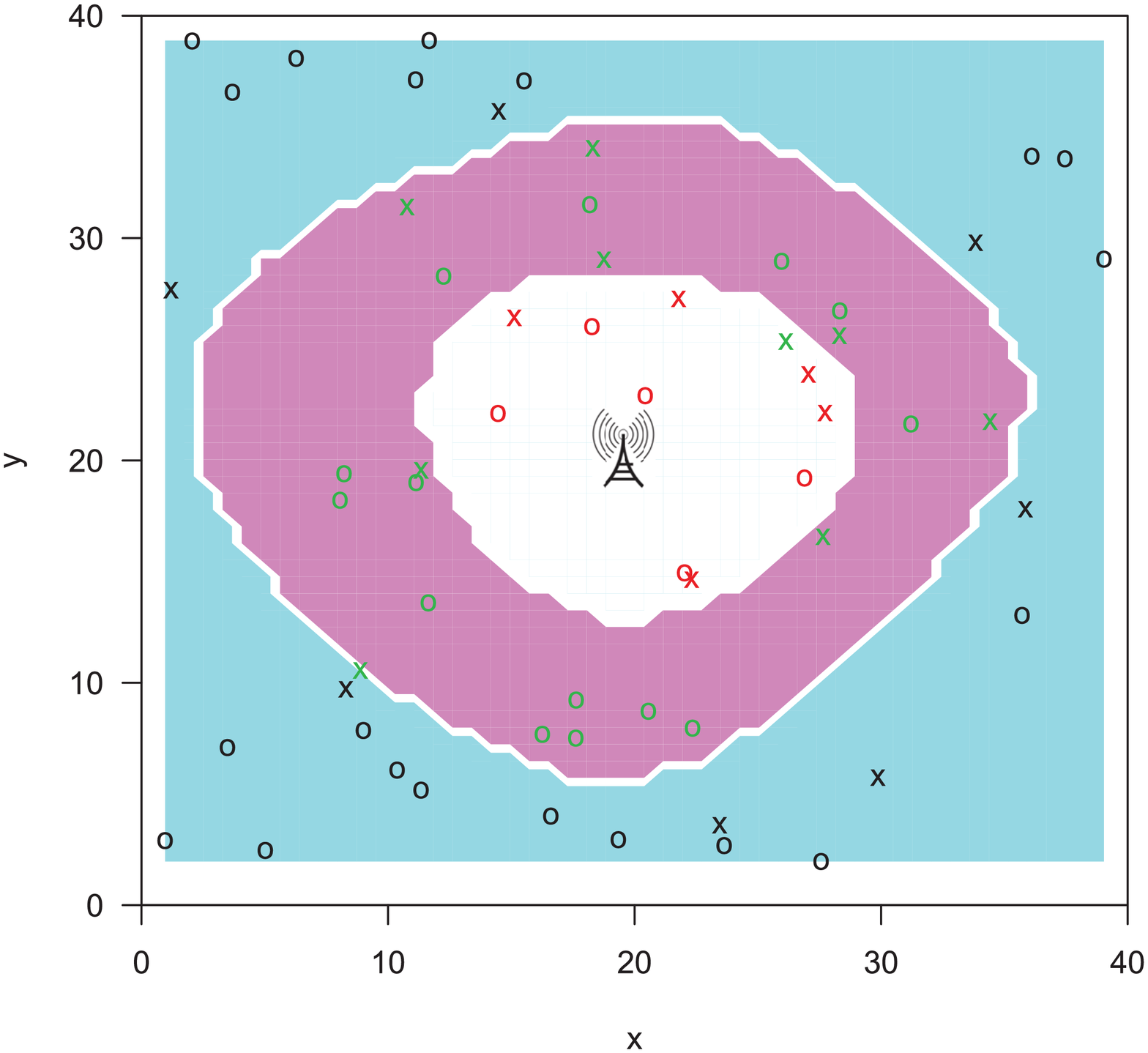}}
   \hspace{-3cm}\subfigure[Band 3, All sensing]{\includegraphics[width=1.9in]{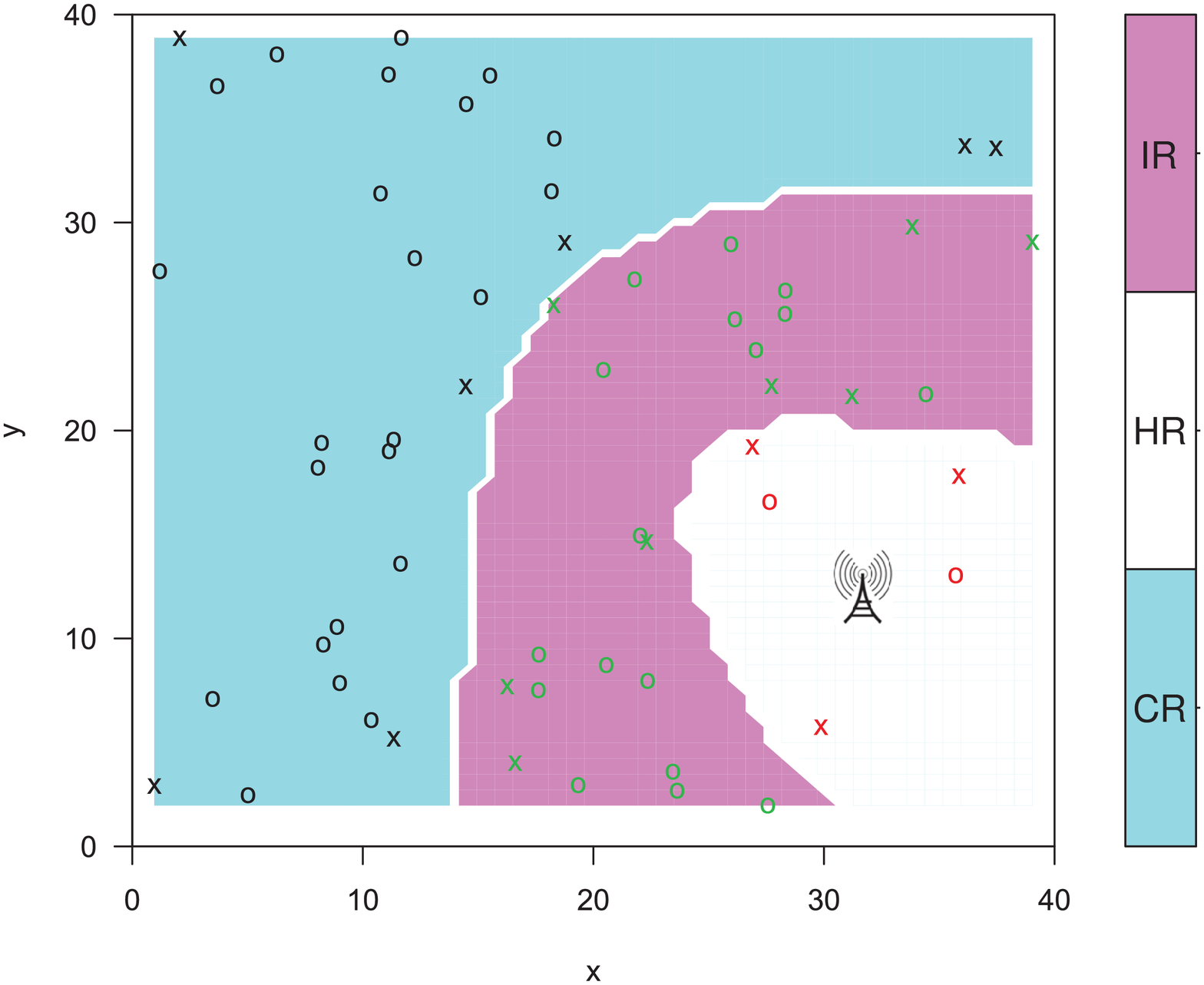}}
   \subfigure[Band 1, 50\% sensing]{\includegraphics[width=1.9in]{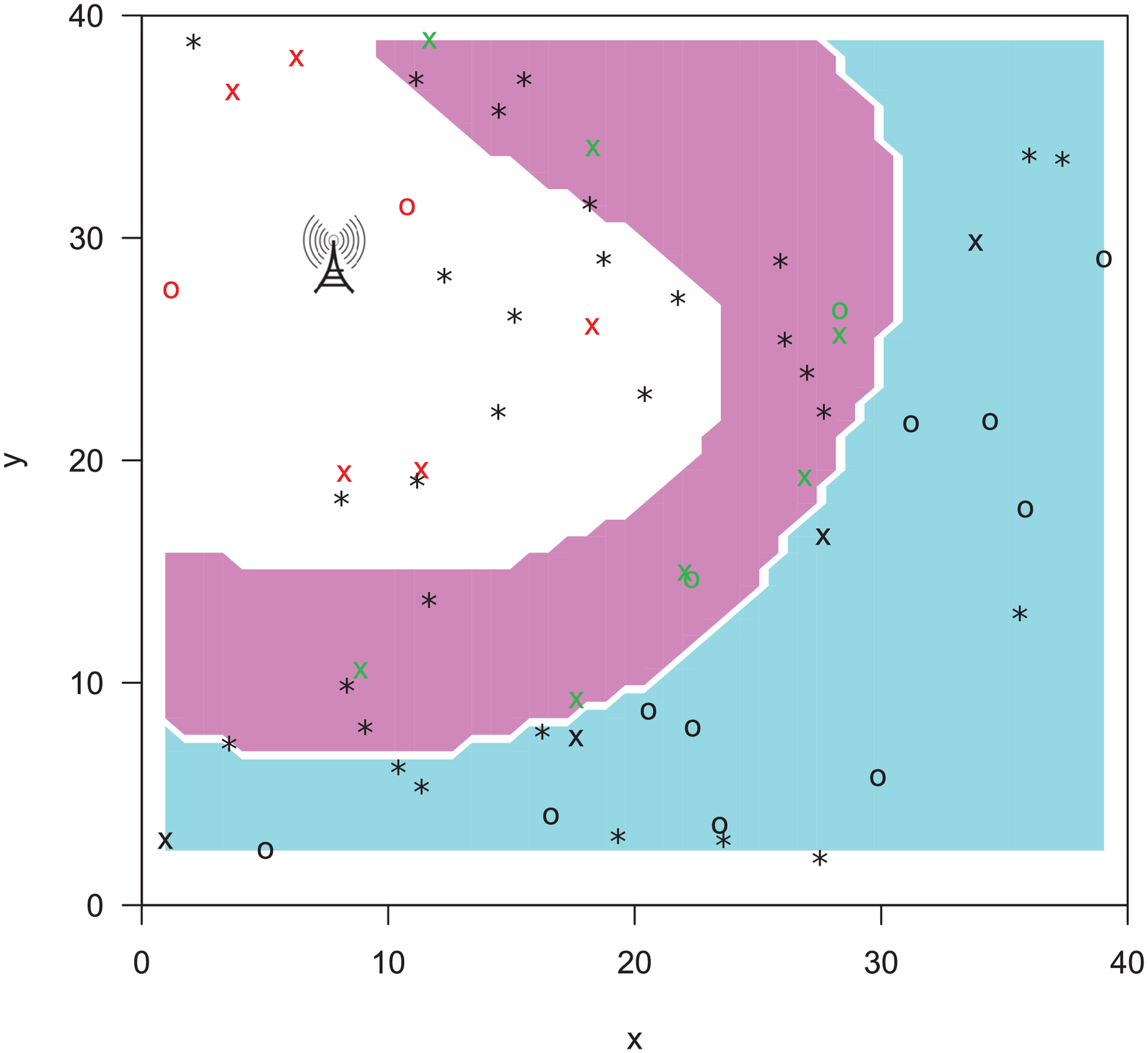}}
   \hspace{-3cm}\subfigure[Band 2, 50\% sensing]{\includegraphics[width=1.9in]{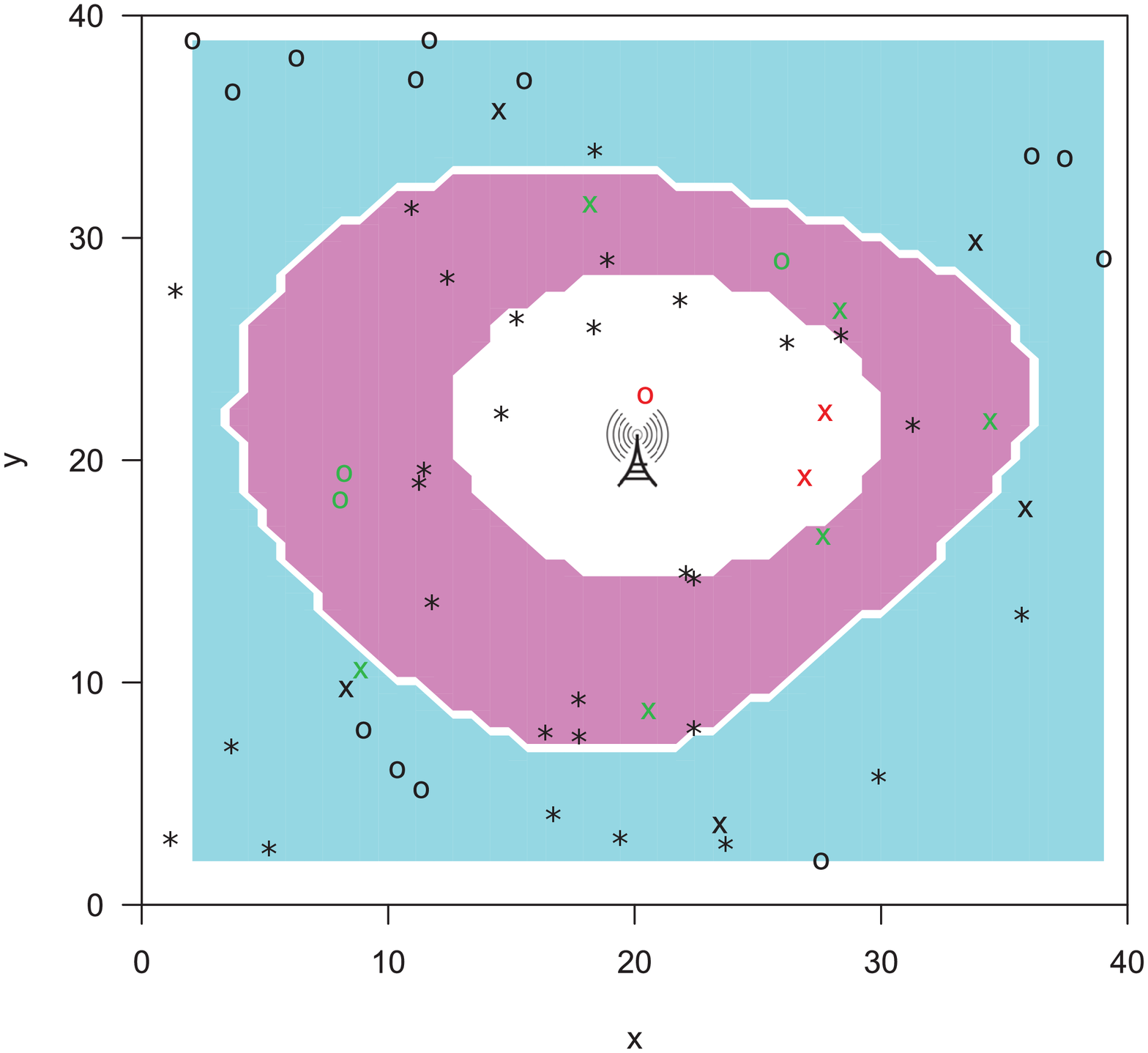}}
   \hspace{-3cm}\subfigure[Band 3, 50\% sensing]{\includegraphics[width=1.9in]{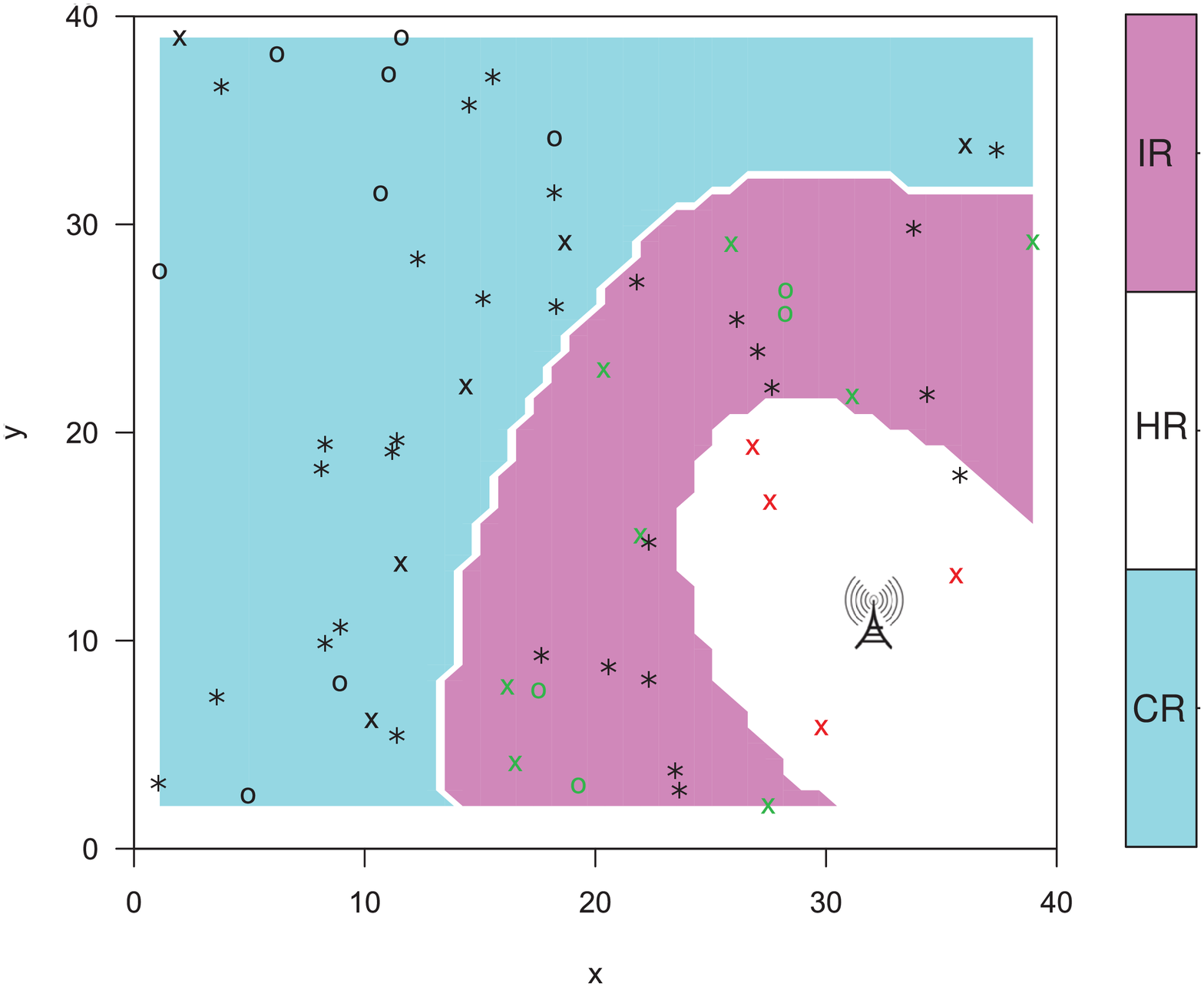}}
\end{subfigmatrix}
\caption{The performance of solution 1 versus solution 2}
\label{tot}
\end{center}
\vspace{1pt}
\end{minipage}
\end{tabular}
\vspace{-15pt}
\end{figure*}
In this section, selected numerical results are provided to investigate the benefits of RF multi-band EH under cognitive radio framework. We consider an area of 40m$\times$40m with seven PUs transmit their signals over seven bands $b=\{0.9, 1.24, 1.56, 1.78, 2.19, 2.46, 2.68\}$ GHz, respectively. Total number of SUs is equal to $M=60$. We assume that the SUs are initially charged with $B_0=P_S T$~mJ of energy. In Table~\ref{tab2}, we present the values of the remaining parameters used in the simulations~\cite{formulas}
{\small
\begin{table}[h]
\centering
\caption{\label{tab2} System parameters}
\addtolength{\tabcolsep}{-2pt}\begin{tabular}{|l|c||l|c||l|c|}
\hline
\textbf{Parameter} & \textbf{Value} & \textbf{Parameter} & \textbf{Value}& \textbf{Parameter} & \textbf{Value}\\ \hline \hline
$T_s$ ($\mu$s) & 1 & $P_s$ (dBm) & 0  &$\eta^b_{m,t}$ & 0.45   \\ \hline
$T$ (s)& $1$ & $P_\text{th}$ (dBm)& $-20$ & $c$ (m/s) & $3\times10^8$ \\ \hline
$\bar{F}$ & 0.1 & $\bar{D}$ & 0.9 & $B_\text{th}$(mJ) & 1 \\ \hline
\end{tabular}
\end{table}}
%

\begin{figure}[h!]
\centerline{\includegraphics[width=3.7in]{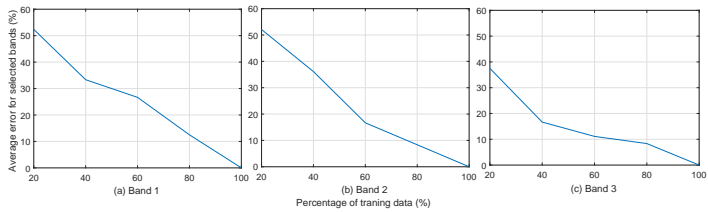}}
  \caption{The effect of choosing training data numbers}\label{error}
\end{figure}

In Fig.~\ref{tot} and Fig.~\ref{error}, because of the paper space limitation, we only select three bands out of seven bands (i.e., $b=\{0.9, 1.78, 2.68\}$ namely band 1, band 2, and band 3, respectively) in order to validate our approach.
Fig.~\ref{tot}, the three regions (HR,IR, and CR) are shown for the two solutions provided in Section~\ref{jsolutions}.
$"x"$ symbol (usually near the boundaries) means that this location of SU plays a significant role in determining the region, while $"o"$ symbol has less significant role in characterizing the regions.
For instance, Fig.~\ref{tot}-a, Fig.~\ref{tot}-b, and Fig.~\ref{tot}-c, show the regions for the three selected bands. It can be deduced that the closest users to the PU can harvest RF energy and the faraway users can transmit their signals without interference with the PU. Also, it can be noticed that the HR for low frequency band (i.e., 0.9 GHz) is greater than the HR for high frequency band (i.e., 2.68 GHz). This is can be justified by knowing that the channel is inversely proportional with the square of the operating frequency. On the other hand, Fig.~\ref{tot}-d, Fig.~\ref{tot}-e, and Fig.~\ref{tot}-f, plot the three regions using random 50\% of SUs as training data.  $"*"$ symbol corresponds to the unselected SUs. Note that, if $"*"$ SUs located in the boundaries, then, it will significantly change characteristics of the the regions. Hence, it will lead to inaccurate regions characteristic.

Fig.~\ref{error} plots the error for the selected bands versus number of training data (i.e., percentage). It is shown that as the training date increased, the accuracy increased. For instant, for band 2 (operated at 1.78 GHz), the error can be reduced from 52\% to around 8\% by using only 20\% instead 80\% of the training data. Therefore, the accuracy can be considered as an important factor in our design. Moreover, this figure, compare between optimal solution: solution 1 (all SUs sensing) when the training data is 100\% and solution 2 (multi SVM) when the training data is less than 100\%.

\begin{figure}[h!]
  \centerline{\includegraphics[width=2.2in]{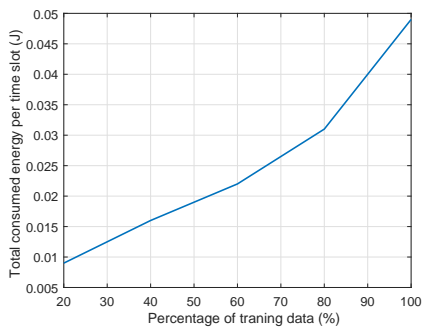}}
   \caption{The effect of training data numbers}\label{energy}
\end{figure}

On the other hand, the total energy consumption is plotted versus the percentage number of training data in Fig.~\ref{energy} to show the trade off of selecting the percentage of the training data. It is obvious to see that as the number of training data increases, the energy consumption increases, this is can be justified by the fact that more training data consume more energy for sensing. For example, the consumed energy is almost doubled by considering 90\% instead of 53\% of the training data. Furthermore, based on optimization problem given in \eqref{ofSVM}-\eqref{c2SVM} one can notice that the complexity increases as we include more training data because it leads to a larger scale problem.
Based on the above discussion, we can design our selection of training data based on three factor; 1- accuracy, 2- energy consumption, and 3- complexity. In other words, given specific range of accuracy, energy consumption limits, and complexity, we can find the optimal number of training data required to satisfy these conditions.

\section{Conclusion}\label{conclusion}

In this paper, we introduced and solved a novel and new optimization problem for multi-band EH schemes under cognitive radio interweave framework.
We proposed a win-win solution, where SUs can sense the spectrum to determine the harvesting and communication geographical
regions, therefore, they can take a decision to harvest or transmit data based on the belonging region.
Also, we jointly optimized number of sensing samples and sensing threshold in order to minimize the sensing energy. A near optimal solution based on GP has been investigated. Furthermore, we have investigated an energy efficient multiple classification approach based on SVM. 

\bibliographystyle{IEEEtran}
\bibliography{References}
\end {document}